# High-Frequency Switching in Superparamagnetic Magnetic Tunnel Junctions by Enhancing Damping

Qi Jia[1,*], and Jian-Ping Wang[1,*]

[1]*Department of Electrical and Computer Engineering, University of Minnesota, Minneapolis, Minnesota 55455, USA*

*Authors to whom correspondence should be addressed:* {jia00096, jpwang}@umn.edu

Abstract: Superparamagnetic magnetic tunnel junctions (sMTJs) are promising components for true random number generation and probabilistic computing. Achieving high-frequency fluctuation while maintaining reliable control over output level is critical for applications. In this work, we systematically investigate the role of magnetic damping in regulating thermal switching rates using macrospin simulations. We show that enhanced damping accelerates the switching rate by increasing the escape rate over the energy barrier. We further compare two control mechanisms: spin-transfer torque (STT) and voltage-controlled exchange coupling (VCEC). Our results reveal that STT-based switching is strongly suppressed under high damping, whereas VCEC, by reshaping the energy landscape without relying on torque-driven dynamics, retains high control efficiency. These findings suggest that enhanced damping not only enables faster stochastic switching in sMTJs but also makes VCEC inherently better suited than STT for high-frequency applications.

Superparamagnetic magnetic tunnel junctions (sMTJs) have attracted increasing attention due to their potential applications in probabilistic computing, stochastic neuromorphic systems, and true random number generators[1–5]. In these applications, the spontaneous switching of the sMTJ between high- and low-resistance states can be electrically detected as a time-resolved signal, from which a random bitstream can be extracted at a fixed sampling rate[6]. However, the lowest usable sampling rate is fundamentally constrained by the intrinsic flipping frequency of the sMTJ[7,8]. For practical high-performance computing, it is therefore highly desirable that this frequency reaches the gigahertz range[9–12].

The switching rate of an sMTJ is proportional to its internal effective magnetic field, and previous efforts to accelerate switching have largely focused on increasing or modifying this field, for instance, by employing in-plane sMTJ structures[7,9,10]. In contrast, relatively little attention has been paid to tuning the magnetic damping constant, which also plays a crucial role in determining the rate. According to Kramers' escape theory, larger damping can accelerate the escape rate over the energy barrier[6,13]. Yet this effect has been largely overlooked in spintronic device design, particularly where spin-transfer torque (STT) and spin–orbit torque (SOT) have been regarded as the primary strategies for local magnetic control[14–18]. In such STT/SOT-driven devices, higher damping increases the critical current and results in greater energy consumption[9,19].

Recently, however, a voltage-driven alternative to STT, voltage-controlled exchange coupling (VCEC), has been demonstrated, which enables bipolar control of sMTJs without relying on damping-like-torque-driven dynamics[3,20–25]. Whether such a mechanism can provide new benefits in the context of sMTJs remains largely unexplored. Here, we address this gap by systematically investigating the role of magnetic damping in stochastic switching using macrospin simulations and by comparing STT- and VCEC-based control.

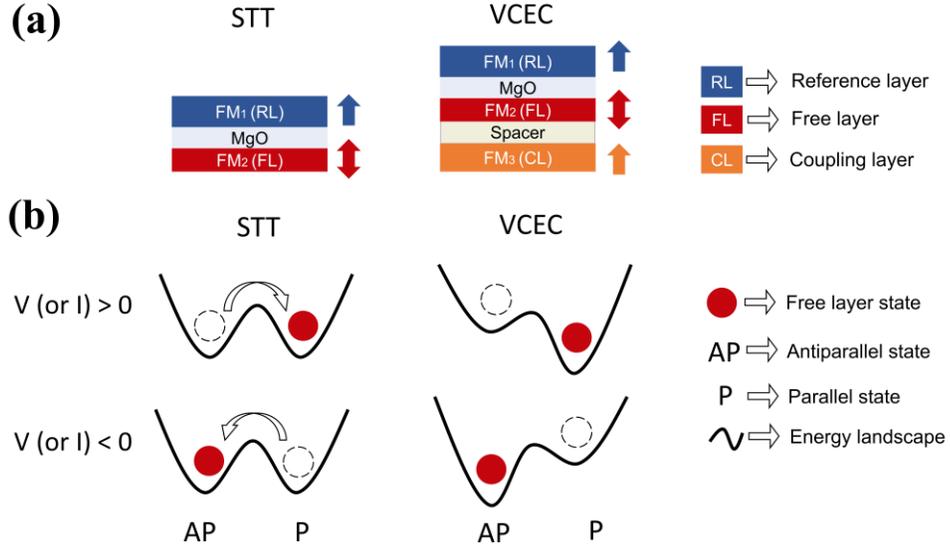

**FIG. 1.** (a) Schematics of the core structure of STT and VCEC stack structure. (b) Qualitative energy landscapes for the two driving mechanisms.

Fig. 1 summarizes the core stack designs for STT and VCEC. In STT stacks, a current through the tunnel barrier polarizes spins in the reference layer; reversing the current reverses the damping-like spin-transfer torque and drives the free-layer magnetization between antiparallel (AP) and parallel (P) states while the underlying double-well energy landscape remains essentially unchanged. In VCEC, an applied voltage across the spacer modulates the interlayer exchange between the free layer and the coupling layer, thereby reshaping the energy landscape, tilting or even inverting the relative depths of the AP and P minima, and enabling deterministic AP↔P switching.

We model the free layer as a single-domain macrospin with a unit vector $m(t) = (m_x, m_y, m_z)$, governed by the stochastic Landau–Lifshitz–Gilbert (LLG) equation in Landau–Lifshitz (LL) form:

$$\frac{d\mathbf{m}}{dt} = -\frac{\gamma}{1+\alpha^2}\mathbf{m}\times\mathbf{H}_{\text{eff}} - \frac{\gamma\alpha}{1+\alpha^2}\mathbf{m}\times(\mathbf{m}\times\mathbf{H}_{\text{eff}}), \quad (1)$$

where $\gamma$ is the gyromagnetic ratio and $\alpha$ is the Gilbert damping. The effective field contains a uniaxial anisotropy along the z direction, an optional VCEC contribution, and thermal noise:

$$\mathbf{H}_{\text{eff}}(t) = H_k m_z \hat{\mathbf{z}} + H_{\text{VCEC}}\hat{\mathbf{z}} + \mathbf{H}_{\text{th}}(t). \quad (2)$$

Thermal fluctuations are modeled as zero-mean Gaussian white noise applied to each Cartesian component, with the standard deviation of $H_{\text{th,std}} = \sqrt{\frac{2\alpha k_B T}{\gamma \mu_0 M_s V \Delta t}}$. Here, $\alpha$ is the Gilbert damping, $k_B$ the Boltzmann constant, T the temperature, $\gamma$ the gyromagnetic ratio, $\mu_0$ the vacuum permeability, $M_s$ the saturation magnetization, V the free-layer volume, and $\Delta t$ the integration time step.

The Slonczewski spin-transfer torque is included as an additive term to Eq. (1) in LL-equivalent form (damping-like component): $\frac{\gamma H_{\text{STT}}}{1+\alpha^2}[\mathbf{m} \times (\mathbf{m} \times \mathbf{p}) - \alpha \mathbf{m} \times \mathbf{p}]$, where $\mathbf{p}$ is the polarization direction and $H_{\text{STT}}$ is the damping-like equivalent field. Unless otherwise specified, simulations use the parameters summarized in Table 1.

Table 1. General Simulation Parameters and Constants.

| Parameters | Symbol | Value |
|---|---|---|
| Gyromagnetic ratio | $\gamma$ | $2.211 \times 10^5$ m/(A·s) |
| Permeability | $\mu_0$ | $4\pi \times 10^{-7}$ T·m/A |
| Gilbert damping | $\alpha$ | 0.2 |
| Boltzmann constant | $k_B$ | $1.380649 \times 10^{-23}$ J/K |
| Saturation magnetization | $M_s$ | $1.0 \times 10^6$ A/m |
| Free layer volume | V | 10nm×10nm×1nm×3.14 |
| Anisotropy field | $H_k$ | $5.0 \times 10^4$ A/m |
| Temperature | T | 300 K |
| Time step | $\Delta t$ | 0.5 ps |
| Total simulation time | $t_{\text{total}}$ | 10 μs |

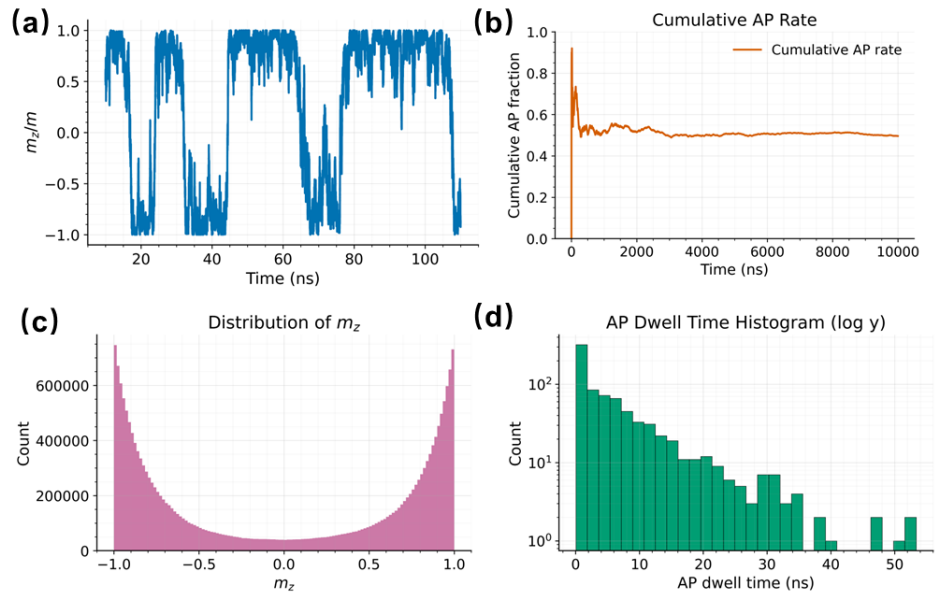

**FIG. 2.** (a) Time evolution of normalized $m_z$. (b) Histogram of the $m_z$. (c) Cumulative antiparallel (AP) fraction versus time. (d) Histogram of AP dwell times (log scale).

Using the parameters and constants listed in Table 1, we performed the macrospin simulation with a fixed Gilbert damping constant of $\alpha = 0.2$. A representative trace of the time-dependent magnetization $m_z(t)$ within a 100 ns time window is shown in Fig. 2(a). Because the free layer possesses a strong out-of-plane magnetic anisotropy, the magnetization preferentially aligns along the ±z directions and therefore spends the majority of time residing in either the AP or P state. When the instantaneous thermally induced random field becomes comparable to or larger than the anisotropy energy barrier, the magnetization undergoes stochastic transitions between the AP and P states, corresponding to thermally activated switching events.

To further illustrate this random switching process, the distribution of $m_z$ over a total simulation time of 10 μs is plotted in Fig. 2(b). The magnetization clearly fluctuates symmetrically around zero, confirming the absence of bias in the system. The antiparallel (AP) rate is defined as the fraction of time during which the magnetization remains in the AP state, labeled as "1". The evolution of the cumulative AP rate as a function of the simulation time is displayed in Fig. 2(c). At short timescales, the rate exhibits strong fluctuations due to the limited number of switching events. As the simulation progresses and more events are accumulated, the rate gradually stabilizes and converges toward 0.5. This result is consistent with the expectation for an unbiased double-well potential, where the two stable states are energetically equivalent in the absence of any external bias field.

The statistical characteristics of the switching dynamics are further analyzed through the histogram of the dwell time shown in Fig. 2(d). When the occurrence frequency is plotted on a logarithmic scale, the distribution approximately follows a straight line, suggesting an exponential decay of dwell time probability. Such behavior agrees well with previous experimental observations and confirms that the thermally activated transition process is well captured by the Néel-Brown model, in which the switching rate follows the Arrhenius-type dependence on the energy barrier and thermal energy.

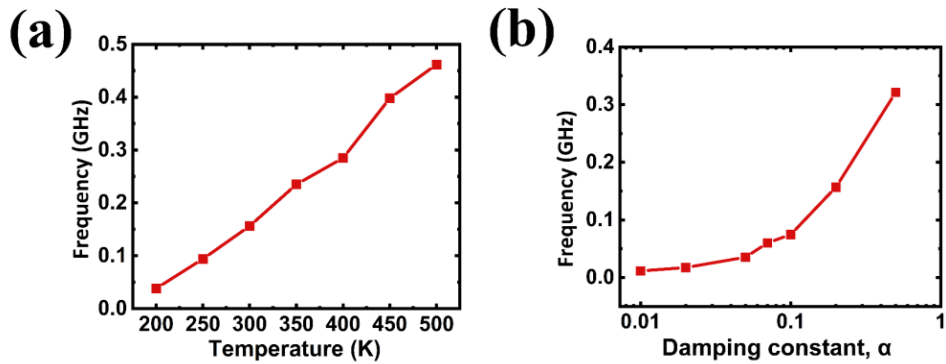

**FIG. 3.** Internal (attempt) frequency of the sMTJ as function of (a) temperature and (b) Gilbert damping constant.

We next varied key parameters to examine their influence on the stochastic switching frequency. First, the damping constant was fixed at α = 0.2, while the temperature was varied. As shown in Fig. 3(a), the switching frequency increases with increasing temperature. This trend is consistent with expectations, since higher thermal energy reduces the stability of the magnetic states and facilitates more frequent barrier crossings.

Subsequently, we investigated the effect of the damping constant while keeping the temperature fixed. The results, presented in Fig. 3(b), show that the switching frequency also increases with increasing damping. This behavior indicates that enhanced damping accelerates the relaxation dynamics, thereby enabling more rapid stochastic reversals. From a device perspective, for applications where high-frequency fluctuations are desirable, such as true random number generation or probabilistic computing, it is advantageous to employ free layers with intrinsically high damping or engineered damping enhancement.

The attempt frequency in the current simulation is relatively low because we focus on perpendicular MTJs. This choice is motivated by their superior compatibility with foundry fabrication processes and the clearer physical interpretation they provide. In contrast, in-plane MTJs can exhibit much higher fluctuation frequencies [9,10], and the method of increasing the magnetic damping can intrinsically elevate the attempt frequency beyond 1 GHz in such systems.

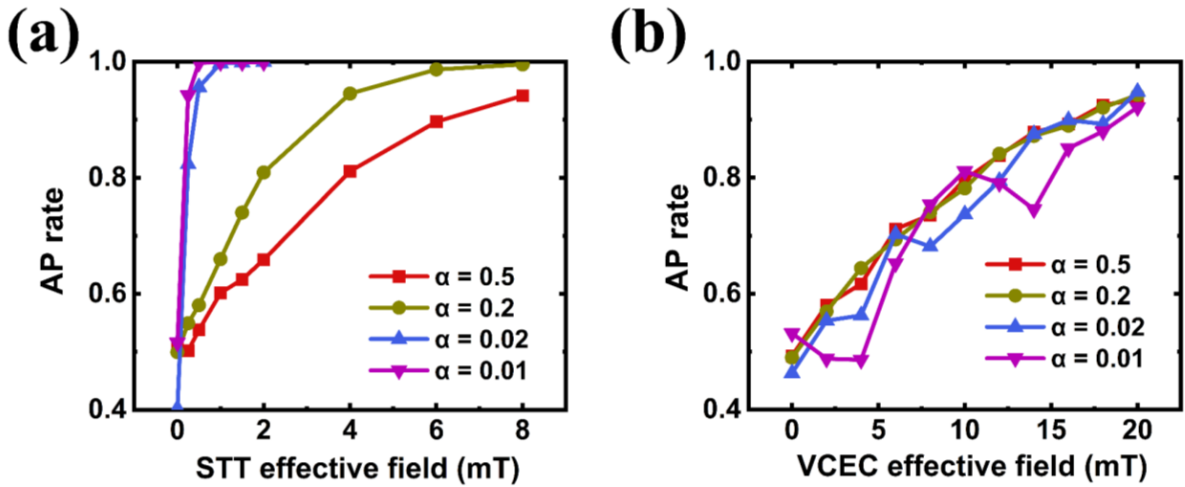

However, this design principle contrasts with that of conventional MTJs controlled by STT, where high damping is generally unfavorable. We simulated the STT case under different damping constants to examine this dependence. As shown in Fig. 4, the effective field manipulates the sMTJ much more efficiently when the damping is small. As the damping increases, a larger effective field is required to tune the antiparallel AP rate to a desired level, indicating reduced torque efficiency.

**FIG. 4.** Comparison of switching efficiency versus the damping constant $\alpha$: (a) spin-transfer torque (STT) and (b) voltage-controlled exchange coupling (VCEC).

In contrast, for the VCEC-driven case, the switching efficiency remains nearly unaffected by changes in damping. This behavior is consistent with the understanding that VCEC acts on the sMTJ through an effective exchange field, rather than through a torque that must overcome damping-related energy loss. Consequently, the VCEC mechanism provides a more robust and energy-efficient control pathway under high-damping conditions.

We also note that the variance of the AP rate becomes larger at smaller damping values. This arises because, under low damping, fewer thermally activated switching events occur within a fixed observation window, resulting in greater statistical fluctuation in the measured rate.

In conclusion, we have systematically investigated the effect of damping and temperature on the stochastic switching dynamics of superparamagnetic MTJs using a macrospin model. The results show that the switching frequency increases with both temperature and damping, in agreement with thermal activation theory. More importantly, the comparison between STT and VCEC reveals a fundamental distinction in their damping dependence. While STT efficiency rapidly deteriorates with increasing damping due to its torque-mediated nature, VCEC remains largely unaffected, as it modifies the magnetic energy landscape directly through an effective exchange field. This robustness against damping not only enables high-frequency stochastic operation but also suggests a pathway toward low-power, voltage-driven spintronic devices. These findings highlight the intrinsic advantage of VCEC over conventional current-induced switching mechanisms for next-generation probabilistic and neuromorphic spintronic applications.

The authors have no conflicts to disclose.

The data that support the findings of this study are available from the corresponding author upon reasonable request.

### Acknowledgement

This material is based upon work supported in part by the National Science Foundation 2230963, ASCENT: TUNA: TUnable Randomness for Natural Computing, and seed grant of NSF Minnesota MRSEC center. This work was supported in part by ASCENT, one of six centers in JUMP, a Semiconductor Research Corporation (SRC) program sponsored by DARPA.